\documentclass[preprint, 5p, twocolumn]{elsarticle}

\usepackage{amssymb}
\usepackage{amsmath}
\usepackage{graphicx}
\usepackage{hyperref}
\usepackage{xcolor}

\journal{arXiv}

\begin{document}

\begin{frontmatter}



\title{A Hybrid Quantum-Classical Particle-in-Cell Method for Plasma Simulations}


\author[inst1]{Pratibha Raghupati~Hegde} 
\author[inst1,inst2]{Paolo~Marcandelli}
\author[inst1]{Yuanchun~He}
\author[inst1]{Luca~Pennati}
\author[inst1]{Jeremy~J.~Williams}
\author[inst1]{Ivy~Peng}
\author[inst1]{Stefano~Markidis}

\affiliation[inst1]{organization={School of Electrical Engineering and Computer Science, KTH Royal Institute of Technology, Stockholm}, country={Sweden},
}
\affiliation[inst2]{organization={Department of Civil and Environmental Engineering, Politecnico di Milano, Milano}, country= {Italy}
}
\begin{abstract}
We present a hybrid quantum-classical electrostatic Particle-in-Cell (PIC) method, where the electrostatic field Poisson solver is implemented on a quantum computer simulator using a hybrid classical-quantum Neural Network (HNN) using data-driven and physics-informed learning approaches. The HNN is trained on classical PIC simulation results and executed via a PennyLane quantum simulator. The remaining computational steps, including particle motion and field interpolation, are performed on a classical system. To evaluate the accuracy and computational cost of this hybrid approach, we test the hybrid quantum-classical electrostatic PIC against the two-stream instability, a standard benchmark in plasma physics. Our results show that the quantum Poisson solver achieves comparable accuracy to classical methods. It also provides insights into the feasibility of using quantum computing and HNNs for plasma simulations. We also discuss the computational overhead associated with current quantum computer simulators, showing the challenges and potential advantages of hybrid quantum-classical numerical methods.  
\end{abstract}




\end{frontmatter}



\section{Introduction}
Plasma, often referred to as the fourth state of matter, consists of ionized gas and composes the majority of visible matter in the universe. The study of plasma is therefore crucial in astrophysics and space science. Additionally, the study of plasma is critical for fusion energy research, accelerator physics, and industrial applications. Several techniques to simulate plasma have been developed, among them the Particle-in-Cell (PIC) method~\cite{birdsall2018plasma}, due to its versatility and robustness, is one of the most employed. In PIC simulations, following a kinetic approach, the distribution function of the plasma particles is modeled by means of macroparticles (or computational particles), which evolve in the phase-space and time according to self-consistent electromagnetic fields. Despite being a powerful tool, the PIC method demands huge computational resources, with the update of particle position and the calculation of fields being the most computationally expensive steps.

Plasma PIC simulations are currently performed on High-Performance Computers (HPCs) using powerful codes such as VPIC~\cite{bowers2009vpic}, iPIC3D~\cite{markidis2010ipic3d}, Warp-X~\cite{vay2018warp}, PIConGPU \cite{burau2010picongpu}, and BIT1~\cite{tskhakaya2007bit}. There are ongoing efforts to utilize emerging technologies such as quantum computing to accelerate the simulations or a part of the plasma simulations~\cite{dodin2021applications}. Quantum computing is a post-Moore technology that has been theoretically shown to provide exponential speed-ups in tasks such as finding prime factors of an integer. It is also speculated to provide advantages in optimization problems, quantum chemistry simulations, and machine learning~\cite{hegde2024beyond}. These advantages are attributed to the utilization of quantum mechanical properties such as quantum superposition, entanglement, and interference within the quantum algorithms~\cite{markidis2024quantum}. Developing quantum algorithms for solving PDEs can revolutionize various fields such as plasma physics and fluid dynamics simulations. In this work, we primarily focus on Quantum Neural Networks (QNNs) to solve PDEs. QNNs mainly utilize parametrized quantum circuits similar to classical neural networks~\cite{markidis2023programming}, and have shown to exhibit high expressivity when compared to classical networks~\cite{du2020expr-pwer-qcircuits}.

\begin{figure*}[t!]
    \centering
    \includegraphics[width=\linewidth]{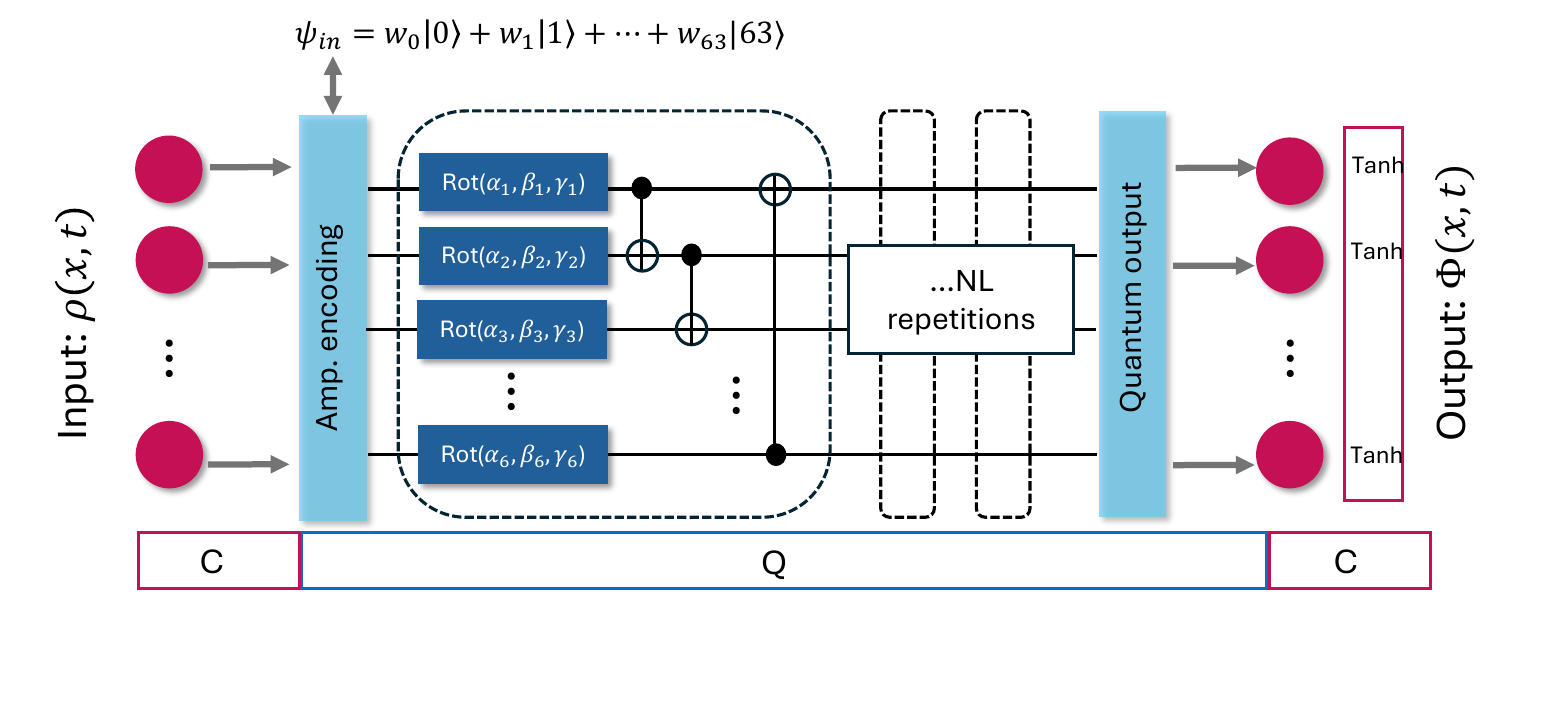}
    \caption{The CQC (classical- quantum -classical) model that is trained to solve the Poisson equation with the charge density $ \rho_t$ at the time step $t$ as the input and the electric potential $\Phi_t$ as the output. }
    \label{fig:Hybrid model}
\end{figure*}

In this work, we propose a hybrid quantum-classical electrostatic PIC method that integrates a quantum Poisson solver within the classical PIC framework. Our main contributions are as follows:

\begin{itemize}
    \item We implement a hybrid classical-Quantum Neural Network (HNN) to solve the electrostatic Poisson equation, replacing traditional classical solvers. The HNN is trained on classical PIC simulation results and executed using the PennyLane quantum computing framework.
    \item We use a physics-informed neural network approach with a Poisson PDE loss to improve the accuracy of predictions, which works particularly well in regimes with sparse data availability.
    \item We integrate the quantum Poisson solver into a classical PIC workflow. The field interpolation and particle motion computations remain on a classical system. The electrostatic field is obtained using a HNN method.
    \item We verify our hybrid approach by applying it to the two-stream instability problem, a benchmark test in plasma physics. We compare accuracy against classical solvers and analyze the behavior of the plasma dynamics.
    \item We evaluate the computational cost of quantum circuit simulations, analyzing execution times and scalability challenges associated with near-term quantum hardware.

\end{itemize}

Our findings suggest that HNN can serve as efficient surrogates for classical Poisson solvers, making it a viable approach for quantum-accelerated plasma simulations.

\section{Background}

In the PIC method, each macroparticle represents millions or even billions of real particles, such as electrons or ions. While particle velocities and position are treated continuously, the actual spatial domain and time are discretized. The space discretization is done by means of a grid, and it is required in order to compute the self-consistent fields generated by the particles. At each PIC cycle, the simulation advances by one time step. Each computational cycle consists of four main steps: firstly, particle quantities such as charge densities and current densities are deposited on the grid with an interpolation process (particle-to-grid interpolation), then, using the interpolated quantities, fields are computed. After that, field values are interpolated back to particle positions (grid-to-particle interpolation), and eventually, the particles are advanced in the phase-space using the interpolated field values with the Newton equations of motion (particle mover). 
In the limit of the electrostatic case, only particle charge densities ($\rho$) are interpolated to the grid, and the self-consistent field is calculated by solving the Poisson equation for the potential $\Phi$:

\begin{equation}
    \nabla ^2 \mathrm{\Phi} = -\rho,
    \label{eq:poisson}
\end{equation}
further, the electric field ($E$) is computed by taking the gradient of the potential:
\begin{equation}
    E = -\nabla \mathrm{\Phi}.
\end{equation}
The PDE in Eq.~\eqref{eq:poisson} is traditionally solved using finite-difference methods. In this work, we use a hybrid quantum-classical neural network to solve Poisson's equation by training it with the data of some of the instances of the baseline PIC iterations. We propose a methodology to offload the task of solving the Poisson equation (Eq.~\eqref{eq:poisson}) to a classical-quantum hybrid neural networks that essentially utilize quantum computers.

\begin{figure*}[t]
    \centering
    \includegraphics[width=0.7\linewidth]{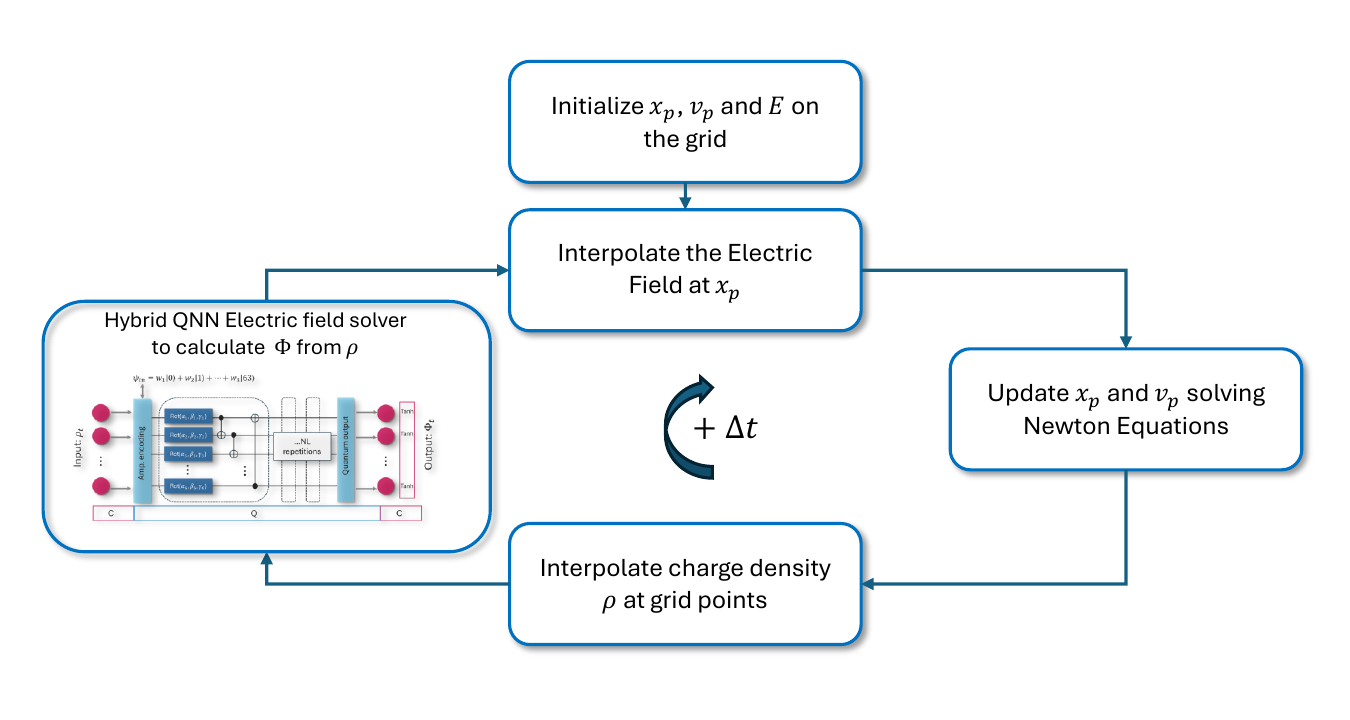}
    \caption{A quantum-classical hybrid PIC method}
    \label{fig:pic}
\end{figure*}
Solving PDEs is an important part of physics simulations and requires huge computational resources in most cases. Developing quantum algorithms that exclusively take advantage of quantum mechanical properties to solve PDEs is one of the rigorously researched topics in recent times~\cite{variational-pde}. Moreover, the current quantum hardware has several limitations when it comes to error correction, qubit scaling, and coherence times, which makes it difficult to implement theoretically successful algorithms such as HHL algorithm~\cite{harrow2009quantum}, which was proposed to solve the system of linear equations. The current quantum computers, also known as Noisy-Intermediate Scale Quantum (NISQ) computers, can run variational quantum algorithms that treat quantum circuits as an ansatz whose parameters are optimized to minimize a cost function. There are several quantum algorithms designed in this framework, such as Quantum Physics Informed Neural Networks (QPINNs)~\cite{kyriienko-qpinns,markidis2022physics,hegde2024quantum}, Quantum Reservoir computing models~\cite{pfeffer2022hybrid}, and Quantum optimization algorithms that can be used to solve PDEs~\cite{qannealing-pde}. 

\section{Related Work}
Several quantum computing methods have been proposed to solve the governing equations of plasma. Refs.~\cite{joseph2023quantum,dodin2021applications,koukoutsis2023quantum} presents an overview of the different strategies plasma simulation in the context of fusion energy and plasma physics, including Hamiltonian simulation approaches and quantum lattice algorithms (QLA)~\cite{koukoutsis2024quantum}. 

Quantum algorithms, in particular linear solvers, have also been proposed to solve discretized PDEs, such as the Poisson equation \cite{wang2024quantum, wang2020quantum2}. Most of these approaches fall within the category of fault-tolerant quantum algorithms. Our work is based on the variational approaches using quantum neural networks (QNNs). The properties of QNNs have been studied in a number of different works, focusing on expressivity, trainability, and power of generalization~\cite{du2020expr-pwer-qcircuits,schuld:dataencoding,du:qcircuit_learnability}.They have been employed to solve problems within varying fields, ranging from computational fluid dynamics~\cite{sedykh2024hybrid, kyriienko-qpinns, pfeffer2022hybrid} to condensed matter physics~\cite{qnn:condensedmatter}.

QNNs have also been used for solving PDEs, for example, through physics-informed neural networks~\cite{kyriienko-qpinns,markidis2022physics, jaderbeg:qpinn}. Previous works, such as Refs. \cite{variational-poisson,markidis2022physics,hegde2024quantum}, focus on using QNNs for solving the Poisson equation. In these works, quantum neural networks were trained to solve the given instance of the Poisson equation.  Integrating such unsupervised methods within the PIC loop can be computationally expensive, as they require a long time to train. In this paper, we use a supervised learning approach, where we train HNNs to solve the Poisson equation that can also generalize to unseen instances. A similar strategy was also adopted in Ref.~\cite{aguilar2021deep}, where the authors introduced supervised deep neural networks and deep convolutional networks to solve the Poisson equation within PIC codes. Moreover, in several of the recent works~\cite{trahan2024quantum,Mari2019TransferLI,sedykh2024hybrid}, classical-quantum hybrid neural networks have been utilized to perform machine learning tasks, which have shown advantages over their classical counterparts when it comes to accuracy, most likely due to the high expressive power of the quantum circuits~\cite{du2020expr-pwer-qcircuits}. Motivated by these recent works, we integrate the HNNs within traditional PIC codes as the Poisson equation solvers.

\begin{figure*}[t]
    \centering
    \includegraphics[width=0.48\linewidth]{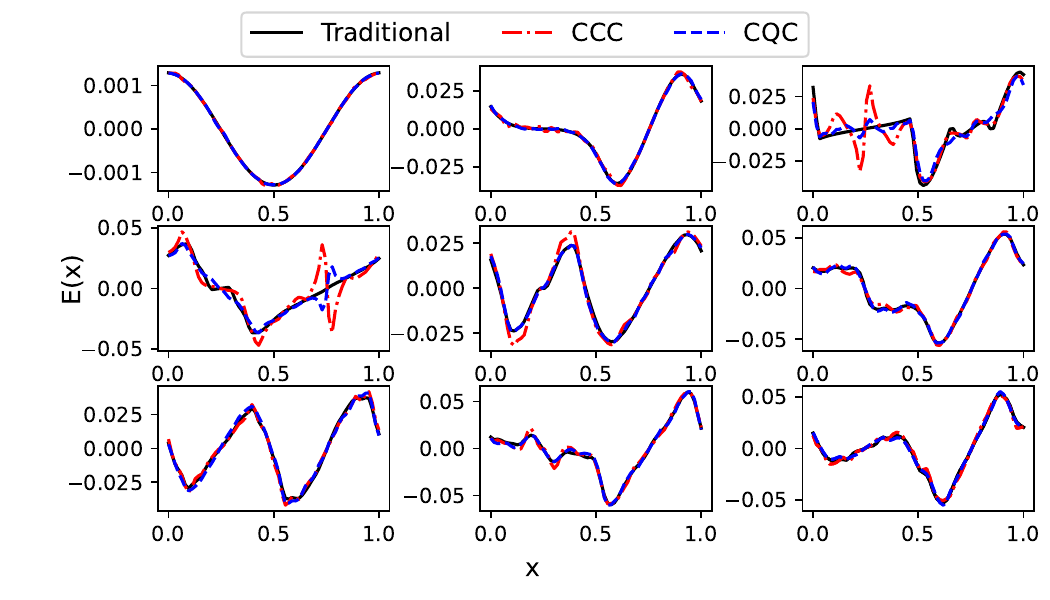}
    \includegraphics[width=0.2\linewidth]{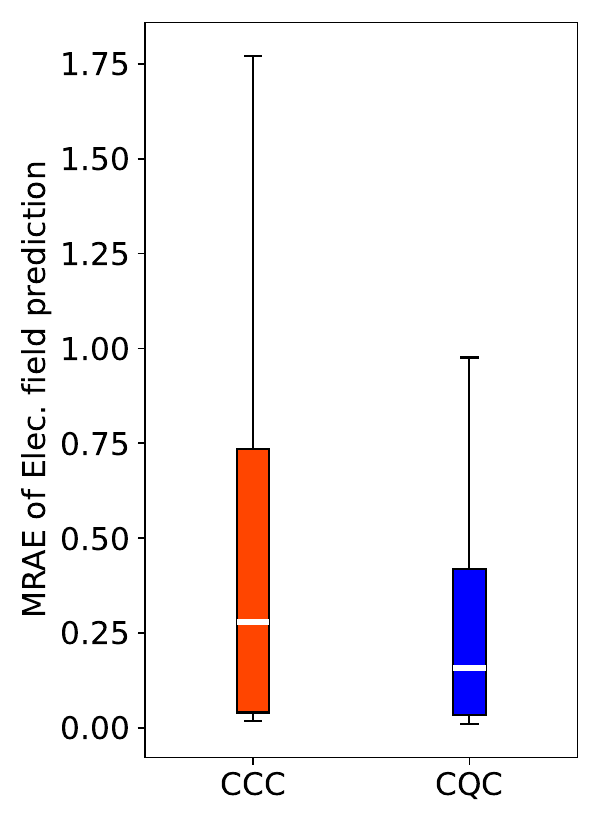}
    \includegraphics[width=0.3\linewidth]{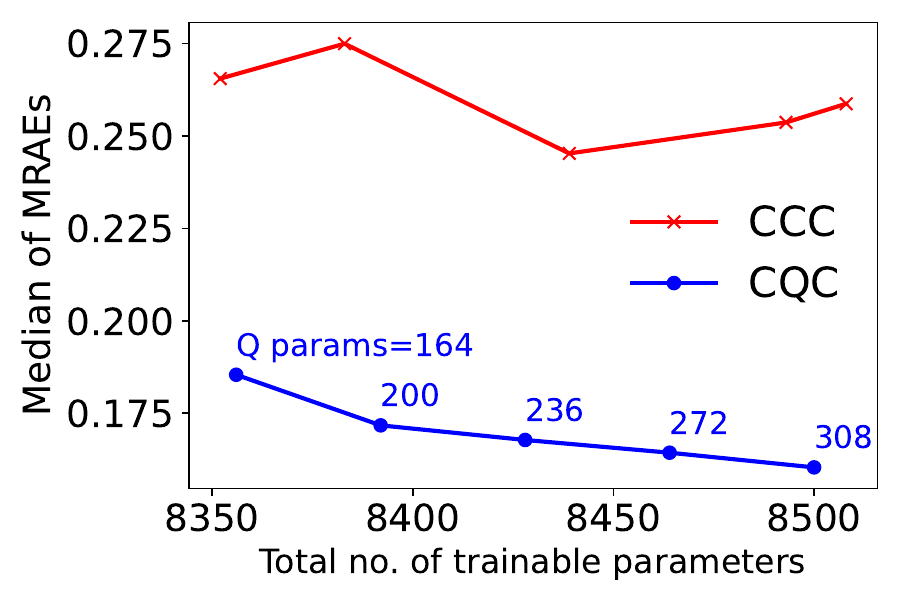}
    \caption{The left figure shows the electric fields $\mathrm{E} = -\mathrm{\nabla \Phi}$ prediction for some of the time-steps of the PIC simulation for $v_0=0.07$. The predictions by the CQC, and CCC PIC codes are compared with the traditional baseline PIC code. The midddle figure shows the mean relative absolute error (MRAE) in predicting the electric fields for 1000 time steps. The right figure shows the median of MRAEs of predicting the electric fields by the HNN models with varying number of quantum trainable parameters (Q params in the figure). The performance is compared with the CCC models with a similar number of trainable parameters.}
    \label{fig:predicts}
\end{figure*}
\section{Methodology}

\subsection{Hybrid Quantum-Classical PIC Implementation.}  

The Hybrid classical-quantum PIC iterations are implemented as shown in Fig.~\ref{fig:pic}, where the traditional electric field solvers are replaced by a supervised hybrid quantum neural network model. The electric field interpolation and particle mover steps are done on a classical computer. 
For the training part, we resort to two techniques: The first one is the purely data-driven approach, where we train the HNN with the electric potential $\Phi (x, t)$ values on every point on the grid for the input source function $\rho (x, t)$ for different time steps of $t$. The second approach is where we train the hybrid network to minimize a physics-informed loss that corresponds to the residual of the Poisson equation for the given input source function $\rho$.

\subsection{Experimental Setup}\label{sec:experimental setup}
The two-stream instability problem that we use to train the model and validate the results has the following parameters. The simulation is initialized with two counter-streaming beams of electrons at velocities $v_0$ (red points in phase-space) and $-v_0$ (blue points in phase-space). We consider the maximum value of $v_0=0.1$ for experimentation. This initial velocity is sufficient to result in the instability within the plasma system. The thermal velocity $v_{th}$ is set to 0. The particles move within a domain of length $L=1$. The PIC simulation set-up has 64 grid points, and each point represents 200 particles. The total simulation time is 50, which is split into 1000 steps (step size = 0.05), and in each step, the PIC cycle is performed to update the particle movement. All the physical quantities discussed in the paper are dimensionless, with the electron charge to mass ratio equal to 1.


\subsubsection*{Hybrid neural network structure}
We train a classical-quantum hybrid neural network to be the Poisson solver that predicts the electric potentials given the charge densities as the inputs. We consider two models of neural networks for our experiments. The first one is the hybrid model with classical-quantum-classical layers (CQC model), and the second is the classical neural network with 3 classical layers (CCC model). The quantum layers are implemented using Pennylane quantum simulator~\cite{bergholm2018pennylane}, where we have used the feature of turning quantum nodes into a torch layer to integrate classical layers available in Pytorch.

The classical layer that we use is always a linear layer (implemented with \verb|torch.nn.Linear|) with input and output features set to 64, as we need to encode data at 64 grid points. The classical layer is always followed by an \verb|ReLU| activation function layer (except for the last layer, where we use Tanh activation function). 

The quantum layer is implemented using a quantum circuit. This layer essentially has three major components as shown in Fig. \ref{fig:Hybrid model}: a data encoding part (sometimes also referred to as a quantum feature map), a variational ansatz block and measurements of the quantum circuit outputs. Here we briefly explain the implementation of each of these crucial parts.

\paragraph{Quantum Embedding}
We use amplitude encoding scheme to embed the incoming data from the classical layer into the quantum circuit. In this encoding scheme, $2^n$ features are encoded in the probability amplitude of $n$ qubit state (As $n$ qubits are described by a vector of dimension $2^n$ in the Hilbert space).  Therefore, in our model, the real number outputs from the first classical layer (of dimensionality 64) are encoded using 6 qubits. The quantum circuit is initialized in the state,
\begin{equation}
    \psi_{in} = w_0 |0\rangle + w_1 |1\rangle + ... + w_{63} |63\rangle,
\end{equation}
where the probability amplitudes are normalized, $\sum\limits_i |w_i | ^2 =1$. We implement this encoding using Pennylane's \verb|qml.AmplitudeEmebdding| tool. Here, the quantum circuit is prepared in the desired initial state using a sequence of uniformly controlled rotations as proposed by Möttönen et.al. \cite{mottonen2004transformationquantumstatesusing}. This scheme requires $2^{n+2}-4n -4$ CNOT gates and $2^{n+2} -5$ single qubit rotation gates for the state preparation.

\paragraph{Variational ansatz}
This is the trainable part of the quantum circuit that generally consists of the rotation gates and controlled-Unitary gates. In this work, we use Pennylane's Strongly entangling layers (\verb|qml.StronglyEntanglingLayer|) template as the variational ansatz. It consists of general rotation (Rot) gate that takes 3 arguments, which are trainable. This is followed by several Controlled NOT (CNOT) gates. The role of CNOT gate is to entangle different qubits. This particular variational ansatz has been used extensively in the literature \cite{schuld:dataencoding,trahan2024quantum,zaman:vansatz,bowles2024betterclassicalsubtleart} and has shown to yield accurate results. We have used the number of trainable layers $\mathrm{NL}=6$ to present our main results. Further, we have tested the usage of other two ansatz layers that are defined in Pennylane. One is \verb|qml.BasicEntanglerLayer| which consists of single parameter rotation gates (RX gates) and CNOT gates \cite{schuld:dataencoding}. The other ansatz is the simplified two-design layers proposed in \cite{cerezo2021cost} (\verb|qml.SimplifiedTwoDesign|) which consists of RY gates (a single parameter rotation gate) and entangling CZ gates. We discuss the performance of each variational ansatz in detail in Sec. \ref{sec:data-driven}.

\paragraph{Measurements}
The qubits undergo a unitary transformation determined by the variational ansatz described previously and the final state of the qubits is subsequently measured. We have experimented with different read-outs of the quantum circuit, which are passed to the following classical neural network layers. The measurement outcome of the quantum circuit can be the quantum state itself. Although the statevector simulators can return quantum state description, on a real quantum computer, advanced quantum state tomography techniques have to be used \cite{bisio:qtomogrpahy}. Alternatively, one can compute the Born rule probability distribution of measurements in the computational basis and in practice this requires reinitializing and measuring the same quantum circuit in multiple shots.  One can also compute the expectation value of an observable from such probability distributions, which turns out to be the most commonly used quantum output in variational quantum algorithms. After comparing the different strategies discussed above, we found that using computational basis measurement probability distributions,  which is also easier to implement in practice, resulted in better accuracy.

\begin{figure}[t]
    \centering
    \includegraphics[width=\linewidth]{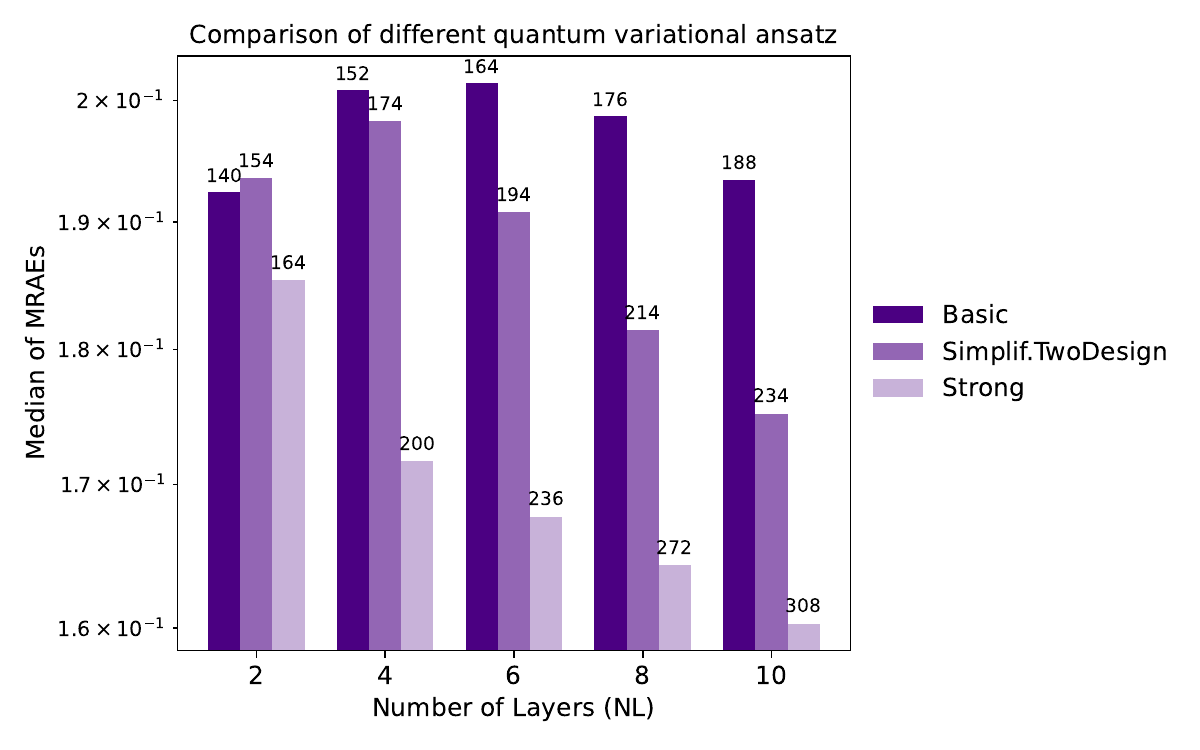}
    \caption{Comparison of the performance of different numbers of layers for the three variational ansatz in HNNs: Basic Entangling layer, Simplified Two-design and Strongly entangling layers. The number of trainable Quantum parameters in each model considered is mentioned on top of the bars.}
    \label{fig:ansatz}
\end{figure}

\begin{figure*}[t]
    \centering
    \includegraphics[width=0.4\linewidth]{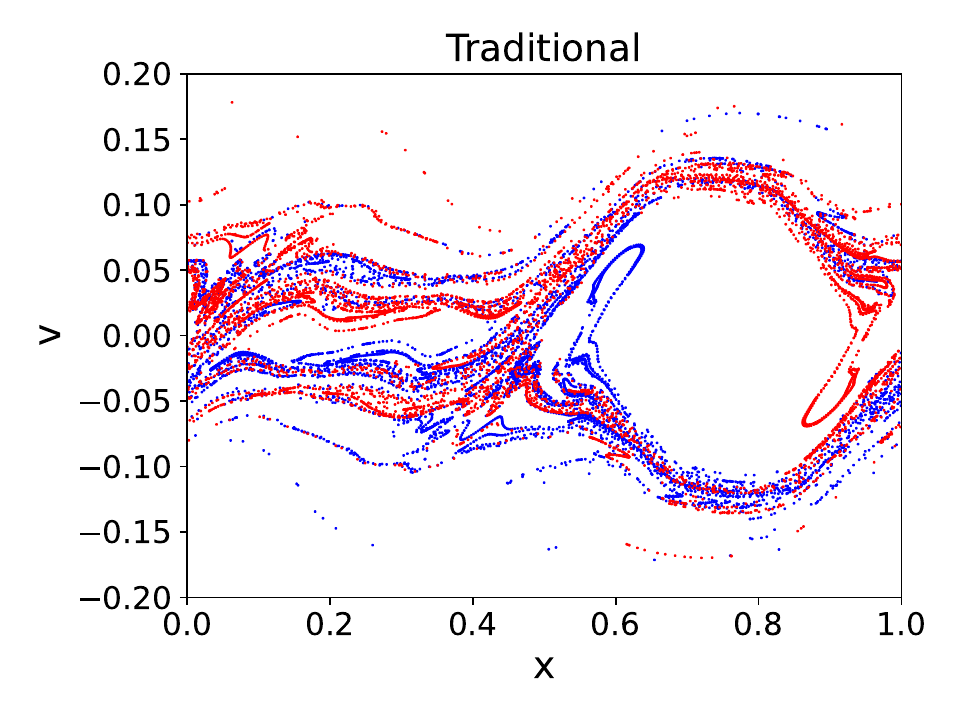}
    \includegraphics[width=0.4\linewidth]{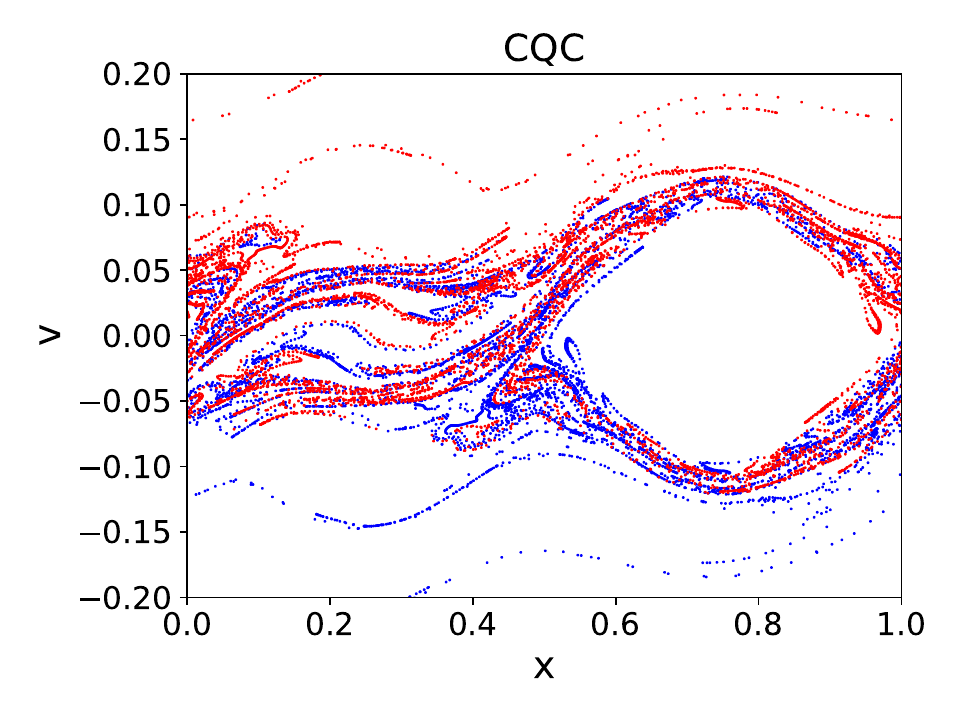}
    \includegraphics[width=0.4\linewidth]{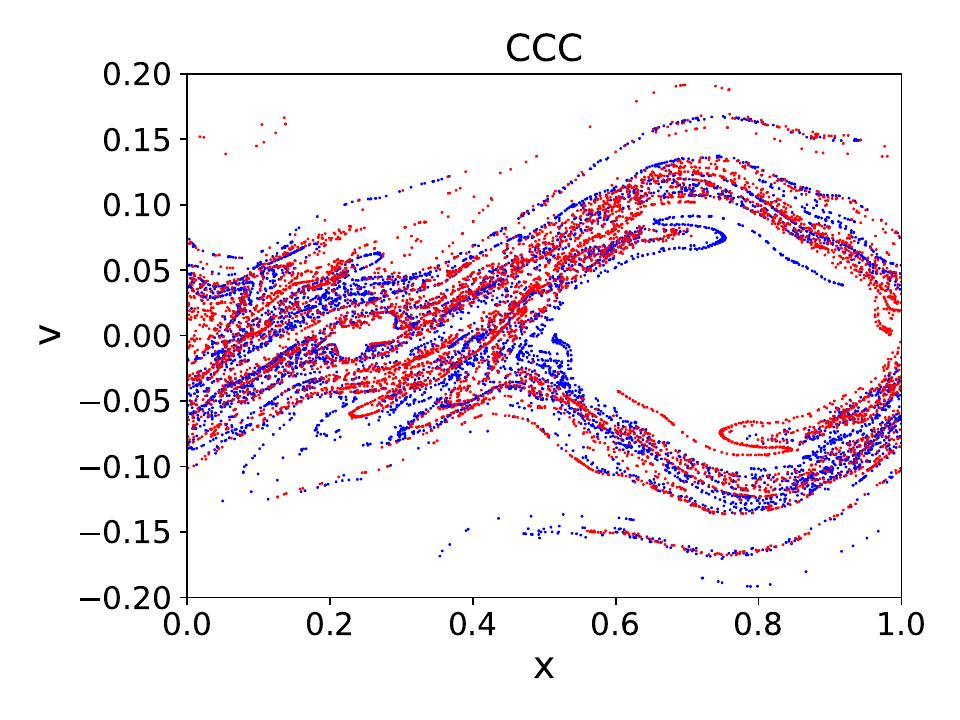}
    \includegraphics[width=0.4\linewidth]{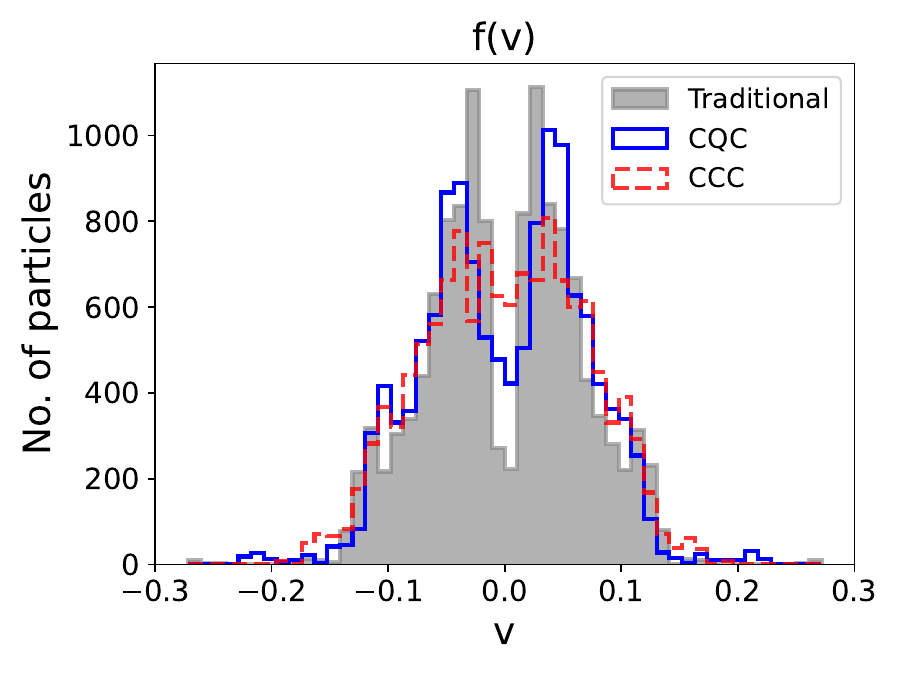}  
    \caption{Phase-space of the particles at the final time of simulation using traditional methods, hybrid neural network (CQC), and classical neural network (CCC). The histogram shows the velocity distribution of the particles at the end of the simulation for the three PIC codes considered.}
    \label{fig:phase-space}
\end{figure*}

\subsubsection{Training}
We simulate the 1D 2-stream instability problem with the traditional linear equation solvers to compute electric potentials for 1000 time steps. We use 500 instances from this dataset as the training data. The HNN receives the charge density at 64 grid-cell points as the input, and there are 64 features in the output.  The output should approximate the electric potential at the 64 grid points when the parameters of the model are optimized using an appropriate loss function.

We use Mean Absolute Error (MAE) between the predicted potential $\Phi_{pred}$ and the true potential $\Phi_{true}$ at $N_g$ grid points as the loss function to train our model in the data-driven approach,

\begin{equation}
    \mathcal{L}_\mathrm{data} = \frac{\sum\limits_i^{N_g} \left |\Phi(x_i, t)^{true}-\Phi(x_i, t)^{pred}\right |}{N_g},
    \label{eq:data_loss}
\end{equation}
which is also subsequently averaged over different time steps considered in the training dataset.
Given that the goal is to solve the Poisson equation, we have also investigated the usage of a Physics-informed Hybrid neural network that is trained with the loss function composed of a PDE loss (residual of the Poisson equation) and a data loss. The physics-informed loss function  used for training is,

\begin{equation}
\begin{aligned}
\mathcal{L}_\mathrm{PINN} &=\frac{\sum\limits_i^{N_g}\left|\frac{\partial\Phi^2(x_i, t)}{\partial x^2} + \rho (x_i, t) \right|}{N_g} \\
& + \lambda \frac{\sum\limits_i^{N_d}|\Phi(x_i, t)^{true} -\Phi(x_i, t)^{pred}|}{N_d},  
\end{aligned}
\label{eq:pinn_loss}
\end{equation}

where $\lambda$ is the relative weight between data loss and physics loss. $N_d$ is the number of data points used to compute the data-loss in the PINN approach. The gradients are computed using \verb|torch.gradient()| module.
 
We have used the Adam optimizer with a learning rate of 0.001 for all our experiments.
We have also preprocessed the inputs ($\rho$) and output dataset ($\Phi$), by simply dividing by the maximum of absolute values (maximum taken over $x$) at each time step,
 \begin{equation}
     \rho(x, t) = \frac{\rho' (x, t)}{ \max\limits_{\{x\}}  |\rho '(x, t)|} ,\,\,\,
     \Phi(x, t) = \frac{\Phi '(x, t)}{ \max\limits_{\{x\}}  |\Phi' (x, t)|}.
 \end{equation}

 When the hybrid neural networks are called within the PIC loop, the charge density is scaled and used as input. The output of the neural network is scaled back using the scaling of the true solutions. The maximum values of $\rho$ can also be used to scale the $\Phi$ values, as they are related by the Poisson equation and the scaling follow a similar behavior.

\section{Results}
This section outlines the results of using classical-quantum hybrid models as Poisson solvers in the PIC simulation of the 2-stream instability test. In particular, we make a comparative analysis of using classical neural networks and classical-quantum hybrid neural networks (CCC and CQC models) as the Poisson solvers within the traditional baseline PIC method. 

\begin{figure}[t]
    \centering
    \includegraphics[width=\linewidth]{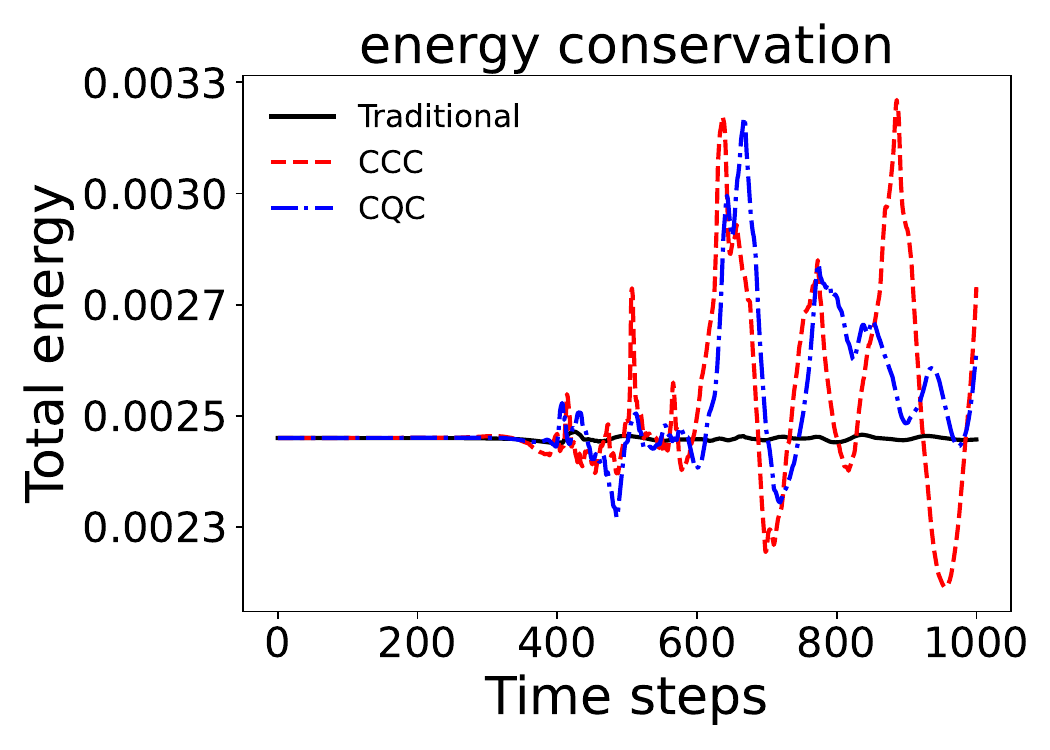}
    \includegraphics[width=\linewidth]{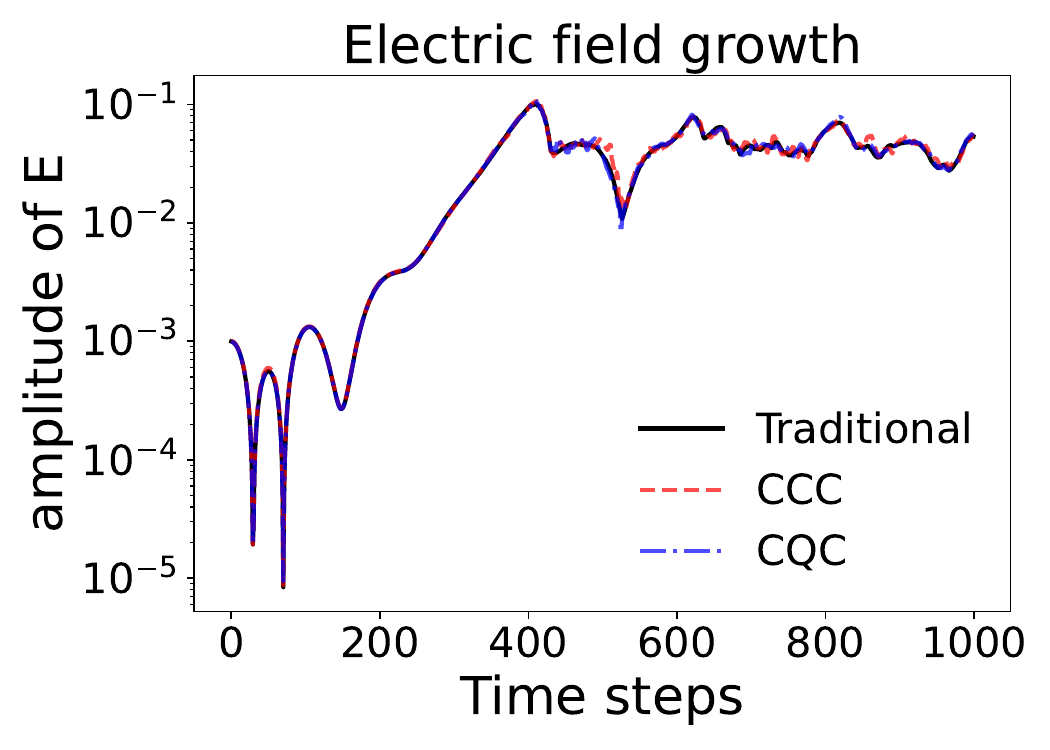}
    \caption{The top plot shows the total energy of the electrons during the PIC simulation. The plot below shows the growth of the electric field amplitude induced by the electrons.}
    \label{fig:energy-eamplitude}
\end{figure}

\subsection{Data-driven supervised learning approach}\label{sec:data-driven}
Here we present the results of using the supervised data-driven approach, i.e, we use the loss function in Eq.~\eqref{eq:data_loss} evaluated at the $N_g=64$ grid points. The training data is generated from a baseline PIC code with initial stream velocities $v_0 = [0.03, 0.05, 0.1]$. Here, the CQC model used has 8,428 trainable parameters,
and the CCC model has 12,480 parameters. Both models are trained for 2000 epochs. The tests are made on the PIC simulation with a new unseen stream velocity $v_0 = 0.07$, which should reveal the generalizing capacity of the neural networks under consideration. 

In Figure~\ref{fig:predicts}, we present the electric field prediction results for different PIC iterations. The left figure shows the electric fields $E(x)$ predictions for some of the time steps when compared with the ones obtained using traditional methods ($\Phi_{exact}$). We observe that the hybrid CQC model can fit the data well, while the CCC prediction deviates from the correct solutions in some of the instances. The plots also demonstrate the highly non-linear nature of the electric fields in question and the diversity in their shapes, which the HNN is able to capture efficiently. We compare the overall predictions for the 1000 time steps of the simulations in the boxplot shown on the right in Fig~\ref{fig:predicts}. The plot shows the distribution of the mean relative absolute errors, MRAE (the relative metric is chosen because the electric field amplitudes lie within a large range of values), where the error values by a CQC model has a lower median value ($\approx 0.168$) and the smaller outliers than the classical CCC model (median $\approx 0.265$). Furthermore, in Fig. \ref{fig:predicts} we evaluate the HNN model using different numbers of variational layers (Strongly entangling layers), each corresponding to a different count of trainable parameters, and report the median of MRAEs for electric field predictions. Our results show a decreasing trend in prediction error as the number of quantum parameters increases. We also compare the HNN's performance to that of the CCC model with a similar number of trainable parameters and observe that, within this range, the HNN consistently outperforms the classical model.

Additionally, we evaluate the performance of various variational ansatz described in Sec.\ref{sec:experimental setup}. Specifically, we consider three circuit structures: Strongly Entangling Layers, Basic Entangling Layers, and a simplified Two-Design ansatz. For each ansatz, we vary the number of layers, $NL = [2, 4, 6, 8, 10]$, and present the corresponding results in Fig.\ref{fig:ansatz}. Among the three, the Strongly Entangling Layers consistently achieve the best performance. This advantage is likely due to the presence of three-parameter single-qubit rotations in the ansatz, which introduce more quantum parameters for a given number of layers. The number of quantum parameters is indicated above each bar in the plot. The number of classical trainable parameters is fixed at 8192 across all configurations. However, it is not evident to what extent the quantum properties such as entanglement is contributing to the high expressive nature of the quantum circuits and requires further investigation.

We finally integrate the trained HNN with the PIC code and present the HNN-PIC simulation results in Fig.~\ref{fig:phase-space}, where we show the phase space of the particles at the end of the simulation for $v_0 = 0.07$. The instability results in a vortex-like structure in the phase space of the particles, which is characteristic of the 2-stream instability problem. One can see that both CQC and CCC models result in a phase space similar to the baseline PIC simulation. However, the velocity distribution $f(v)$ as shown in the histogram reveals that CQC leads to a phase space that is closer to the traditional PIC simulation. We computed the energy distance (a statistical test to compare the similarity between two probability distributions as described in \cite{rizzo2016energy}, implemented using the python module \verb|scipy.stats.energy_distance|) between the predicted velocity distribution and the baseline PIC simulation results. The lower the energy distance, the more similar the distributions are to each other. The energy distance between CQC PIC and baseline PIC was found to be $0.012$, and that between CCC and baseline PIC was found to be $0.016$. This test confirms that the CQC PIC simulated phase space is closer to the one simulated by the traditional PIC method in terms of the energy distance measure. 

We gain further insights into the problem by comparing the total energy of the plasma throughout simulations, as shown in Fig~\ref{fig:energy-eamplitude}. The total energy of the particles during the simulation should be conserved, and the numerical value should be $\approx 0.0024$. While in the initial phase, both models conserve energy; at later times, the total energy is not conserved. We see that the CQC-PIC code shows an energy fluctuation that is slightly smaller when compared to the energy fluctuations of the CCC-PIC code. In addition, we also track the electric field amplitudes during the evolution. The electric field amplitudes are similar to the traditional PIC simulation, with the exponential growth of the electric field in the initial phase and the saturation of amplitudes when the instability is fully developed \cite{Hou:2streaminstability}.

\begin{figure}[t]
    \centering
    \includegraphics[width=\linewidth]{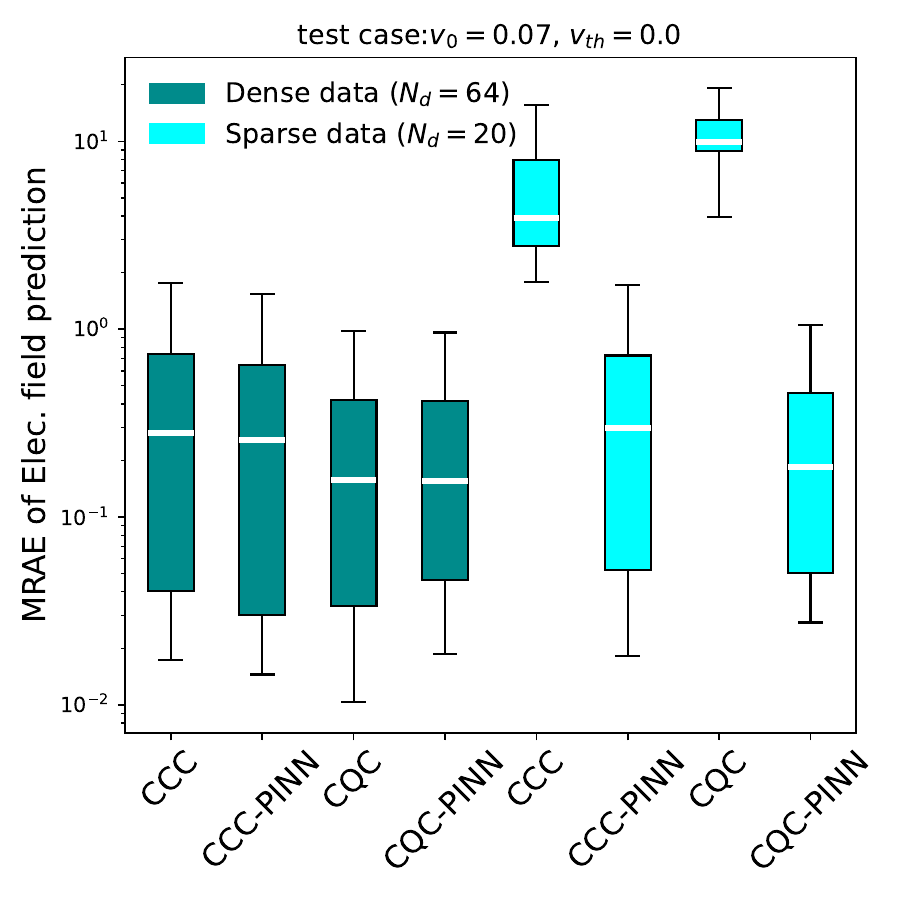}
    \caption{Mean relative errors of electric field prediction for the PIC simulation of the test case $v_0 = 0.07$ using physics-informed approach.}
    \label{fig:pinn}
\end{figure}

\subsection{Physics-informed learning approach}
In this section, we present the results obtained using physics-informed neural networks integrated with PIC code for the 2-stream instability problem. In this case, the neural networks are trained with the loss function $\mathcal{L}_\mathrm{PINN}$ in Eq.~\eqref{eq:pinn_loss}. All the models considered for the experimentation in this section are trained for 3000 epochs.


Physics-informed approaches are known to produce reliable results even in the presence of sparse data. Therefore, here we have considered two scenarios when the training data (electric potential $\Phi$) is dense, in other words, has a higher spatial resolution, and another case when the training data has solutions at sparse points. This test can potentially mimic the situation of low grid resolution data from the experiments or simulations to train the neural networks with. We train the CCC and CQC models with both data-driven and physics-informed approaches. We compare the errors (MRAE) in predicting the electric field ($\mathrm{E} = -\mathrm{\nabla \Phi}$) for the case of $v_0 = 0.07$ when the models are trained with the data for the cases $v_0 = [0.03, 0.05, 0.1]$. 

As the first test, we use $N_d = 64$ data points (same as the data-driven strategy) to compute the data loss $\mathcal{L}_\mathrm{data}$ and the Poisson PDE loss $\mathcal{L}_\mathrm{PINN}$ evaluated at $N_g = 64$ points in Eq. \eqref{eq:pinn_loss}. We use a relative weight $\lambda=0.3$ (for CCC) and $\lambda=0.5$ (for CQC) between data loss and physics loss. In Fig. \ref{fig:pinn}, the first four boxplots show the prediction results with the dense data scenario. We observe a marginal improvement in the CCC model when using the PINN approach. In the case of CQC, PINN results in negligible improvements. This is likely due to the fact that the performance of the neural networks is saturated due to the good quality of data which also obey the correct PDEs. Therefore, the physics-informed approach is not further enhancing the prediction accuracy.

As the second test, we consider sparse points of the electric potential $N_d=20$ in the training dataset. The physics loss is computed for $N_g=64$ collocation points. The prediction results for the case of sparse data are shown in the last four boxplots in Fig. \ref{fig:pinn}. Purely data-driven approaches using both HNNs and classical networks perform poorly in generalising to predict electric fields for the test case. However, we observe a drastic improvement when using PINN approach for both CCC and CQC models in comparison to the respective purely data-driven approaches. Furthermore, sparse-data PINNs perform with errors on par with dense-data models. We have also confirmed that the improvement in the performance by the PINN method is statistically significant by performing Wilcoxon signed-rank test \cite{rey2011wilcoxon} between different pairs of error distributions depicted in Fig. \ref{fig:pinn}, and we have obtained p-values $<0.01$ in each case.

Moreover, we have used the finite difference method (which is implemented using \verb|torch.gradient()|) to compute the gradients within the physics loss as we are dealing with the discrete set of $\Phi$ values of the neural network output. Using more efficient methods, such as automatic differentiation techniques to define $\mathcal{L}_\mathrm{PINN}$, can improve the physics-informed strategy and make it more effective.

\begin{figure}[t]
    \centering
    \includegraphics[width=\linewidth]{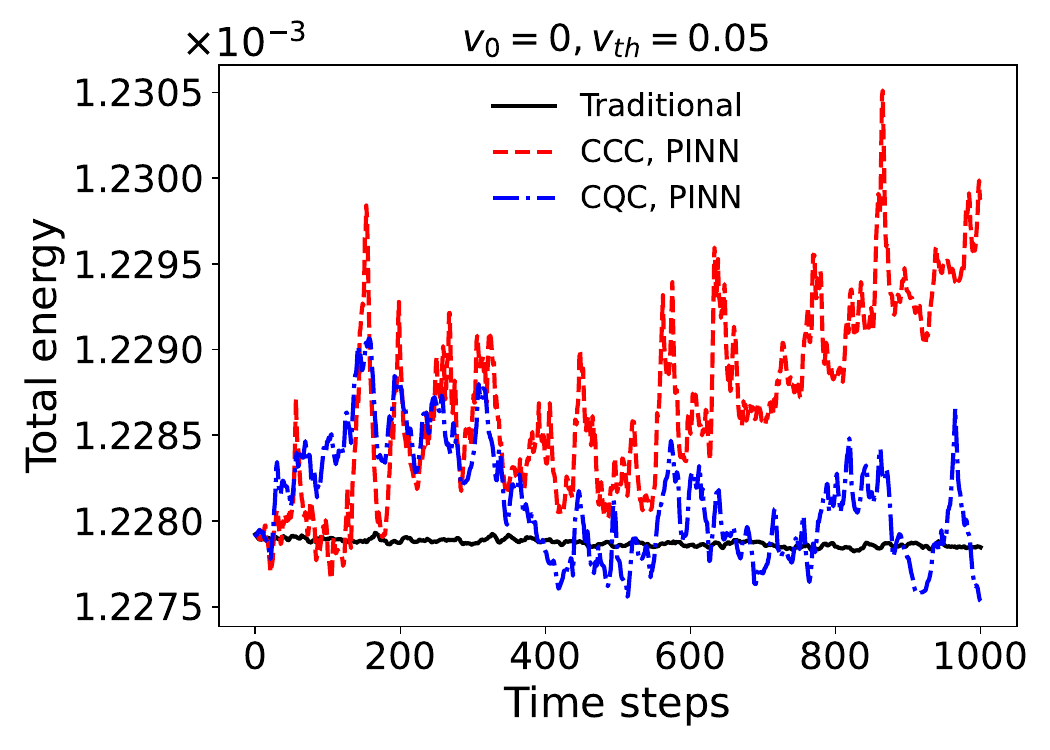}
    \includegraphics[width=\linewidth]{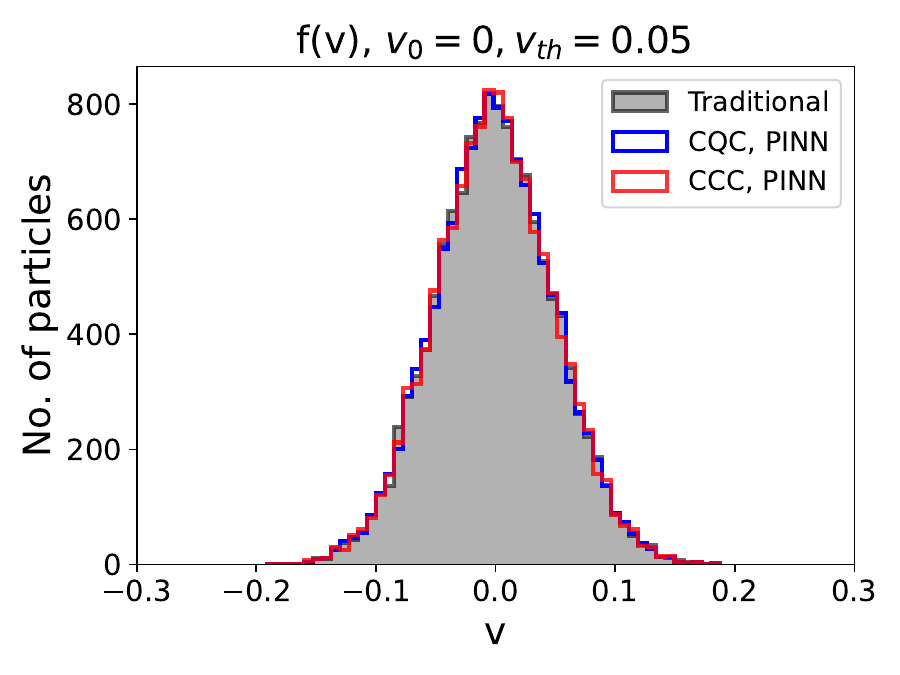}
    \caption{Energy conservation and velocity distribution for a beam of electrons with thermal velocity 0.05. Simulation results compare the prediction by CCC and CQC models using a physics-informed approach.}
    \label{fig:v0_0}
\end{figure}

We test the physics-informed HNN approach for another special case of 2-streams of electrons, where the two beams are initialized with a normal distribution of thermal velocities $v_{th}$ with mean equal to 0. In this case, theoretically, there should not be any growth in the induced electric field, and the plasma should remain stable. The total energy is also conserved in such a system. The CCC and CQC models are trained with the data of $v_{th} = [0.01, 0.03, 0.07, 0.1]$. We make a test hybrid PIC simulation for the case of $v_{th} = 0.05 $. In Fig.~\ref{fig:v0_0}, we show the results of total energy and the final velocity distribution. Both CQC and CCC lead to small energy fluctuations, with CQC showing lower fluctuation than CCC.  We further study the velocity distribution at the end of the simulation in each case. The energy distance computed between CQC-PINN and traditional $f(v)$ is 0.00061, and that between CCC-PINN and traditional $f(v)$ is 0.00059. The energy distance values and the histogram shown in Fig.~\ref{fig:v0_0} reveal that both the classical and hybrid models result in a similar velocity distribution at the end of the simulation and match well with that of the traditional PIC simulation.

\subsection{Parallelization, Performance and Computational Cost}

When it comes to training times, quantum/hybrid neural networks have major drawbacks as they take much longer to simulate. In this work, we adapt an MPI-based data parallelism strategy to enhance training efficiency in the HNN model training process. Specifically, the entire dataset is evenly partitioned and distributed across multiple parallel processes. Each process independently conducts forward and backward propagation on its assigned data subset. Prior to updating the model parameters with the optimizer, gradient synchronization is executed through MPI reduction across all processes, ensuring consistency in parameter updates. Finally, the individual loss values computed by each process are averaged to determine the overall global loss.

\begin{figure}
    \centering
    \includegraphics[width=0.8\linewidth]{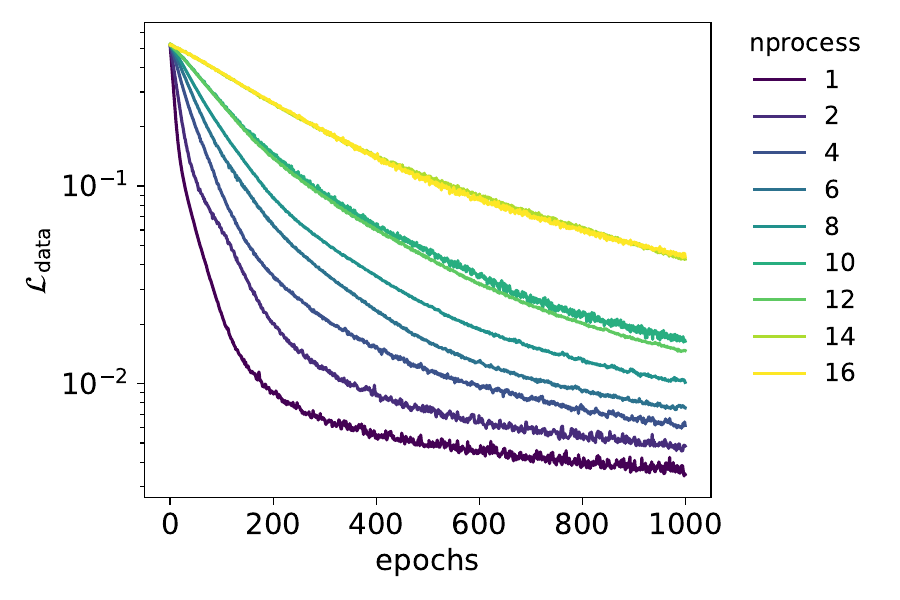}
    \includegraphics[width=0.6\linewidth]{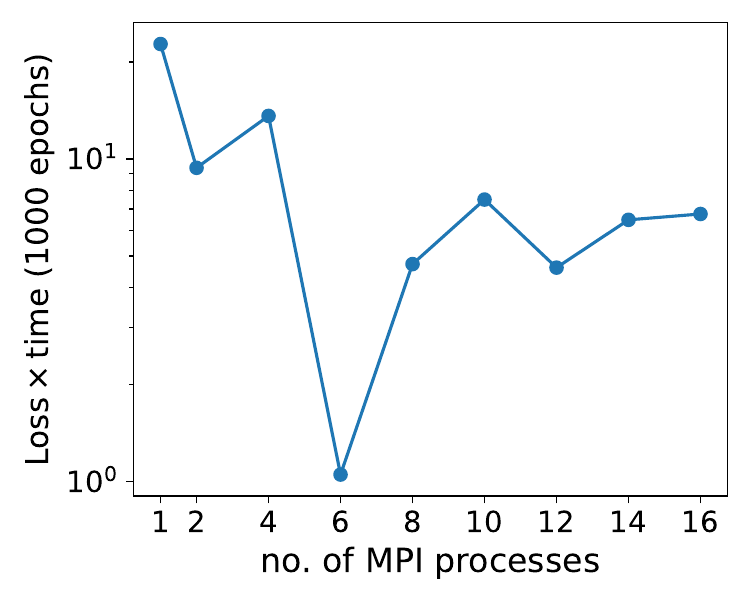}
    \caption{Results of MPI-based data-parallelization strategy. The first plot shows the training of CQC network using different numbers of processes. The second plot shows the trade-off curve between speedup and loss function minimization.}
    \label{fig:nprocess}
\end{figure}

We show the training results of CQC model using multiple MPI processes in Fig.~\ref{fig:nprocess}. The data parallelization helps in reducing the time required to complete a single epoch. However, the averaging of loss leads to slow convergence of the loss function over the number of epochs. In Fig.~\ref{fig:nprocess}, first, we show the loss reduction for 1000 epochs using the MPI processes from 1 to 16. At the end of 1000 epochs, different processes have reached different stages of training. In the second plot, we show the tradeoff in the number of processes and the loss minimization after 1000 steps. The plot indicates that using 6 MPI processes works best in terms of both (among the number of processes we have experimented with). A CQC model takes $\approx 700s$ to complete 1000 epochs (the training data size is 500) of training, while CCC model takes $\approx 37 s$. A PIC simulation with CCC and CQC as Poisson solvers takes $55s$ and $28.5s$ respectively for 1000 PIC iterations. The CQC model has $8,428$ trainable parameters, and the CCC model has $12,480$ parameters. It is to be noted that a CQC model has fewer trainable parameters, which leads to reliable results, which could be a potential advantage when scaled to large models applied to more complicated problems.

\section{Conclusion and Discussion}
In this work, we have proposed a hybrid quantum-classical PIC method for plasma simulations. In particular, we use a supervised quantum-classical hybrid neural network (HNN) to compute the electric potential given the charge densities, in other words, to solve the Poisson equation. We have tested our approach for the simulation of the two-stream instability test, a standard benchmark problem for the PIC codes. We have demonstrated that HNNs can offer a performance advantage in solving Poisson equations—and, by extension, in simulating plasma behavior—with relatively fewer parameters compared to classical models, as shown through multiple experiments. However, it remains an open question how the inherent ``quantumness" of the circuits systematically contributes to this performance gain \cite{bowles2024betterclassicalsubtleart}. A key challenge in applying QNNs to large-scale plasma simulations lies in understanding how performance scales with the dimensionality of the PDEs being solved. Additionally, it is important to investigate the trainability and expressivity of quantum circuits under realistic noise conditions, especially given the lack of efficient error correction protocols on current quantum hardware.

The other challenges of employing data-driven methods for simulating plasma include the availability of the data. The experimental techniques, such as Langmuir probes used to measure the electric potential within plasma, can be invasive and can have low spatial resolution \cite{godyak:probe}. We have shown that the usage of a physics-informed training approach can help overcome this disadvantage for both the classical models and HNNs. Our tests indicate that the physics-informed approach can improve the learning strategies when it comes to predicting electric fields for an unseen charge density distribution, and they are particularly effective in the presence of sparse data.


Additionally, it is also important to investigate the error propagation in electric field prediction within the PIC loop and how the HNNs can be optimized to mitigate such propagation. Moreover, the proposed hybrid PIC method can also be generalized to solve Gauss' law ($\nabla . \mathrm{E} = \rho$) instead of the Poisson equation to obtain the electric field. In the classical electrostatic PIC methods, solving Gauss' law is avoided as the results can be noisy, and it is not straightforward to define boundary conditions. However, in our proposed methods, such as data-driven or physics-informed approaches, we already provide the solutions at several grid points to train the neural network, which should be sufficient to obtain unique solutions of the equation. Further, Gauss' law can be used to define a physics loss for the physics-informed approach.

We have also analyzed the computational cost of training the neural networks. In particular, we have resorted to MPI-based data-parallelization scheme and explored the training performances with the varying number of parallel processes. A CQC model takes about 19x times that of the CCC model to complete the training, showing that long HNN training time on simulators is the major challenge of testing and benchmarking quantum machine learning techniques at present. It could also be a possible direction to utilize accelerators such as GPUs to speed up quantum circuit simulations for the purpose of quantum machine learning, which could also allow us to study the performance scaling in the number of qubits and the variational ansatz depth \cite{cudaquantum}. 

Finally, we note that classical-quantum hybrid algorithms involve rapid exchange of large data between classical and quantum processors. This requires tightly-integrated classical and quantum computing parts, which allow low-latency workflows \cite{hpcqc}, a capability that is currently lacking in quantum systems accessible via the cloud. Achieving tight integration is a challenging task from both hardware and software points of view \cite{unifiedqplatform}. Advances in programming models and software tools aimed at overcoming these challenges \cite{NETZER2026107977,cudaquantum,unifiedqplatform} enhance the feasibility of deploying HNNs on heterogeneous classical–quantum systems. Such capabilities can, in turn, enable their use in computation-intensive tasks such as plasma PIC simulations, which is the topic of interest in this article.


\section*{Acknowledgments}
 Funded by the European Union. This work has received funding from the European High Performance Computing Joint Undertaking (JU) and Sweden, Finland, Germany, Greece, France, Slovenia, Spain, and Czech Republic under grant agreement No 101093261, Plasma-PEPSC (\url{https://plasma-pepsc.eu/}). 
 \newline
 \noindent Partial financial support from ICSC - “National Research Centre in High Performance Computing, Big Data and Quantum Computing”, funded by European Union – NextGenerationEU, is gratefully acknowledged.

\bibliographystyle{elsarticle-num} 
\bibliography{mybibliography}

\end{document}